\documentclass[aps,pre,groupedaddress,amsmath,showpacs, twocolumn]{revtex4} 

\bibpunct{}{}{,}{s}{}{}

\usepackage{dcolumn} \usepackage{graphicx} \usepackage{natbib}

 \newcommand{\figwide}{7.8cm}

\begin{document}

\title{Conditions for building a community of practice in an advanced physics laboratory} 
\date{\today}

\author{Paul W. Irving} 
\author{Eleanor C. Sayre} 
\affiliation{Department of Physics, Kansas State University, Manhattan, Kansas 66506}

\begin{abstract} 
In this paper we explore the theory of communities of practice in the context of a physics college course and in particular the classroom environment of an advanced laboratory. We introduce the idea of elements of a classroom community being able to provide students with the opportunity to have an accelerated trajectory towards being a more central participant of the community of practice of physicists. This opportunity is a result of structural features of the course and a primary instructional choice which result in the development of a learning community with several elements that encourage students to engage in more authentic practices of a physicist. A jump in accountable disciplinary knowledge is also explored as a motivation for enculturation into the community of practice of physicists. In the advanced laboratory what students are being assessed on as counting as physics is significantly different and so they need to assimilate in order to succeed.

\end{abstract}


\maketitle


\section{Introduction}


The development of a professional identity is a fundamental part of student development\cite{Pierrakos2009}. An appropriate subject-specific identity is a strong influence on students' persistence in a discipline \cite{Barton2000, Chinn2002, Cleaves2005, Shanahan2007}. There is a strong relationship between the development of a professional, subject-specific identity and participation in a related community \cite{Hunter2006, Bonnar2007, Lave1991}; in fact, professional identity and community participation are inextricably and symbiotically linked\cite{Lave1991, Del-Castillo2003a, Li2011}.





Laboratory work in particular is generally seen as an opportunity for students to learn problem solving and develop their understanding of physics as well as to understand how the science community works and to eventually be able to take part in the community themselves.\cite{Danielsson2007}. 



In this paper, we claim that structural and programmatic features of a junior-level Advanced Laboratory course (``AdLab'') at Kansas State University, supported by instructor strategies, promote students' enculturation into the physics community of practice by fostering a classroom learning community engaged in bench research.  We support our claims with ethnographic interviews and with observations of AdLab students.

\section{Communities of Practice framework}

We use a communities of practice framework to describe how students develop a classroom community in AdLab.  Communities of practice have three key characteristics: the individuals within form a group, either co-located or distributed\cite{Wenger1998, Coakes2006}; the group has common goals or shared enterprise\cite{Wenger1998, Barab2002}; the group  shares and develops knowledge focused on a common practice \cite{Wenger1998, Barab2002}. This final characteristic can be extended to include the sharing of mutually defined practices, beliefs, values and history.  


An individual participates in several, overlapping communities of practice.  A physics student involved in a research group might also be the goalie on a sports team, for example.  That same research group might be part of a larger collaboration and, at the same time, members of the research group are also members of the physics department.  Because one individual participates in several, overlapping communities, it is important to study how ``more expansive networks''\cite{Nespor1994, Bonnar2007} affect individuals' participation.  Active participants in different communities of practice have opportunities to learn the knowledge, rituals, and histories valued within each community\cite{Charney2007, Hsu2010, Lave1991, Lopatto2009}.  Inasmuch as the different communities overlap, knowledge and practices learned in one community affects practices in another\cite{Li2011, Furman2006, Aschbacher2010}.  Conversely, when communities have different values, individual members may have difficulty importing practices from one community to another\cite{Aikenhead1996, Costa1995}.

In this paper, we are primarily concerned with two overlapping communities: the community which develops within AdLab, and the generalized physics community to which students are aspiring members. These two communities share many goals and norms; AdLab is part of students' training to become physicists; some (but not all) of the practices in AdLab are common to the professional practice of physics.  Of course, the students in these two communities are also members of other communities, but we do not focus on those aspects of their identity in this paper.

\subsection{Duration}

Frequently, communities of practice evolve and grow for extended periods\cite{Wenger1998} and may involve many participants over time.  In these communities, new-comers are socialized into the community of practice through mutual engagement with and support of old-timers.  Through low-level but authentic practices, these peripheral participants are slowly inducted into the knowledge and skills of a particular practice.  Over time, they develop more understanding, knowledge, and skills, becoming central participants and eventually mentoring their own peripheral participants\cite{Lave1991, Barab2002}.   

Students in the process of moving from being a peripheral participant to central participant are referred to as having a trajectory towards being a central member of a community. Being on a trajectory within a community of practice is generally considered a slow induction process \cite{Lave1991}. In the AdLab course students are exposed to a greater number of the authentic practices of members of the community of practice of physicists. We believe that different classroom communities of practice provide different levels of authentic practice and therefore the opportunity for students to accelerate their own trajectory towards becoming a central member of a discipline based community. 

Other communities of practice have shorter duration, such as the length of a semester, and may have fewer members.  Classrooms as communities of practice are well-studied\cite{Borasi1998, Schoenfeld1992, Lemke1990, Berland2011}.  In these shorter-term, temporally-bounded communities\cite{Nathan2005}, we discard the idea of new-comers and old-timers in favor of the more general idea of peripheral and central participants.  Legitimate peripheral participants may sit on the outskirts of classroom discussion, learning discourse and norms\cite{Cobb2001} as they gradually become enculturated\cite{Li2011}.  Conversely, central participants may speak frequently in discussion, be more active in setting norms, or interact with more participants.


\subsection{Learning}

Learning physics is a primary objective in a physics classroom. In a community of practice, learning can conceptualized using situated cognition \cite{Danielsson2007}, participation theory \cite{Rogoff1996, Goertzen2011}, and socially constructed knowledge or understanding \cite{Lampert1990, Lave1991}, or as a process of becoming a member of a community \cite{Hsu2010, Lopatto2009}.  Under these models, learning physics is not merely about learning the contents of physics textbooks, but also about learning ways to participate in the cultural enterprise of professional physicists.  
	
\subsection{Tension between scientist and classroom practices}

If courses like AdLab are to prepare students to be physicists -- to become more central participants in the physicist community of practice -- then those students should engage in legitimate peripheral activities in the physicist community.  Though physics classrooms and the larger physicist community share many of the same norms and practices, they differ in several key respects\cite{Bruffee1993, Cockrell2000, Duschl2002, Squire2003}.  For example, traditional teaching laboratories tend to emphasize reproducing prior results rather than creating new knowledge\cite{Hogan2001}. Introductory physics classes tend to promote students solving many problems weekly while professional physicists work in large teams over multiple years to solve single problems. 

To counteract this disconnect between school science practices and professional ones, the teacher can take on the role of a broker, acting as a go-between the two communities and guiding the classroom community closer to that of the practicing physics community.  She can promote classroom norms and allow activities that are legitimate activities of physicists\cite{Demaree2009}. More advanced coursework is more likely to enact norms and practices that are more like those in the larger professional community, as many faculty are more likely to treat advanced students as junior physicists.

\section{Instructional Context}

At Kansas State University, AdLab is traditionally taken by sophomores and juniors, both physics majors and physics minors.  It meets twice weekly for three hours each meeting; experiments usually take two to three weeks to complete.  Class time is almost entirely devoted to laboratory work, with student presentations once during the semester. The students produce an individual laboratory report for each experiment.  The experiments include common topics in modern physics such as the Lifetime of the $\mu$ meson and Microwave Optics.  Like many upper-level laboratory classes, each experimental set-up has only one set of equipment.  Students rotate through the experiments, and each student will perform a subset of the total number of experiments available.

The advanced laboratory is described as the following in the course catalogue: ``The completion of experiments of current and/or historical interest in contemporary physics.  Students develop skills in and knowledge of measurement techniques using digital and analog instruments.  Various data analysis techniques are used.''


There were 18 students enrolled in the lab at the beginning of the semester and 17 finished the semester; students were organized into six groups. Group members stayed together for the first three experiments and then switched some members for the final three experiments.

\subsection{Structural and Instructional features}
 
Within AdLab, there are several reasons for the development of a classroom community. We find four structural features:
\begin{description}
\item[Paucity of instructor time:]  There are six groups working on six different experiments, each of which is complicated and prone to conceptual, experimental, or equipment difficulties.  There is one instructor.  She simply does not have enough time to spend with each group.  When students need help, they must frequently turn to other sources. 
\item[All in the room together:] All groups work in the same room at the same time.  Because they are in close proximity to each other, there are more chances for interaction between groups.
\item[Experiments long and hard:] The experiments last two or three weeks, and involve many complicated or finicky equipment, difficult error propagation techniques, or conceptual complexity.  This has two implications for community formation: students need to seek out resources to help with troubleshooting their own experiments, and (at any given time) they have time available to help their peers troubleshoot a different experiment.
\item[Same experiments at different times:] Because groups cycle through experiments, pockets of localized expertise develop.  When a new group starts on an experiment, the last group to perform that experiment has direct, localized expertise about performing it.
\end{description}

Additionally, we find one primary instructional choice which supports the development of a community of practice within AdLab.  The instructor of the class, recognizing the structural constraints above, deliberately encourages the sharing and developing of knowledge and understanding between lab groups.  

\subsection{Elements of classroom community}
These four structural features, supported by the instructional choice, work in concert to promote the development of a classroom community of practice. This classroom community of practice has several elements as a result of the structural features and the instructional choice that are not typical of a classroom community. We will refer to these elements as enculturation elements as these elements encourage some of the authentic practices of physicists.

\begin{description}
\item[Classroom norms and expectations:] The students have a greater control over the norms that are negotiated within the classroom. These norms are negotiated over time but result in a more collaborative learning environment and in norms that are more similar to those of professional physicists. The same is true for expectations as students expectations of what counts as physics changes over time due to the jump in ADK.
\item[Distributed expertise:] The students become experts in different experiments which encourages collaboration when groups experience problems with specific experiments.
\item[Community involvement:] The students collaborate and socialize between groups a significant proportion of their time within the AdLab environment.
\item[Many central players:] The socializing and collaboration is not focused on one particular group and is instead distributed throughout all the groups over the length of the AdLab course.
\item[Instructor is not sole mediator:] As the community developed the students began to perceive the instructor as not the sole mediator of learning.
\end{description}

We believe that all four of the structural features are necessary for these enculturation elements to develop. If there were enough instructor time, then students would be more likely to turn to the instructor(s) for help, even if the other three features were present.  If the students were not working in the same room at the same time (as happened in the previous laboratory course), the barriers to intergroup interaction would be larger because students would have to seek each other out outside of class, and they would not have the equipment in front of them as they discussed the experiments.   If the experiments were too simple, the students would not need much help, and if the experiments were too short, they would not have enough time to visit with their colleagues.  Finally, if they all performed the same experiment at the same time, they would all develop expertise at about the same rate, so it would be more difficult for more localized pockets to develop.  Also, if all groups work on the same experiments at the same time, they are likely to develop similar difficulties at similar times, encouraging the instructors to do mini-lectures on specific kinds of troubleshooting and discouraging inter-group discussion.


\section{Methods}

The research presented in this paper is part of a ongoing ethnographic research project on the identity development of undergraduate physics students.  

As a methodology, ethnography originates in anthropology\cite{Pirie1997b, Garfinkel1967}  and is commonly used to understand community life\cite{Marcus1998, Barab2002}.  Ethnography is generally concerned with the sociocultural features of an environment, including how people interact and their discursive practices \cite{Brown2004, Pirie1997b}.  In educational settings, it is used to investigate ``classroom culture'', characterizing various relationships and events\cite{Collins2004, Brown2004}. 


\subsection{Data Sources}

Ethnography typically draws its data from a number of sources in order to get a more complete picture of the culture of the classroom but also in an attempt to overcome some of the weaknesses of subjectivity through triangulating multiple viewpoints \cite{Ernest1997, Emerson1993, Barab2002, Barton2000, Hunter2006, Case2011}. Our data are drawn from diverse sources to triangulate multiple viewpoints on student experiences in Adlab.

\begin{figure} 
\begin{center} 
\includegraphics[width=\figwide]{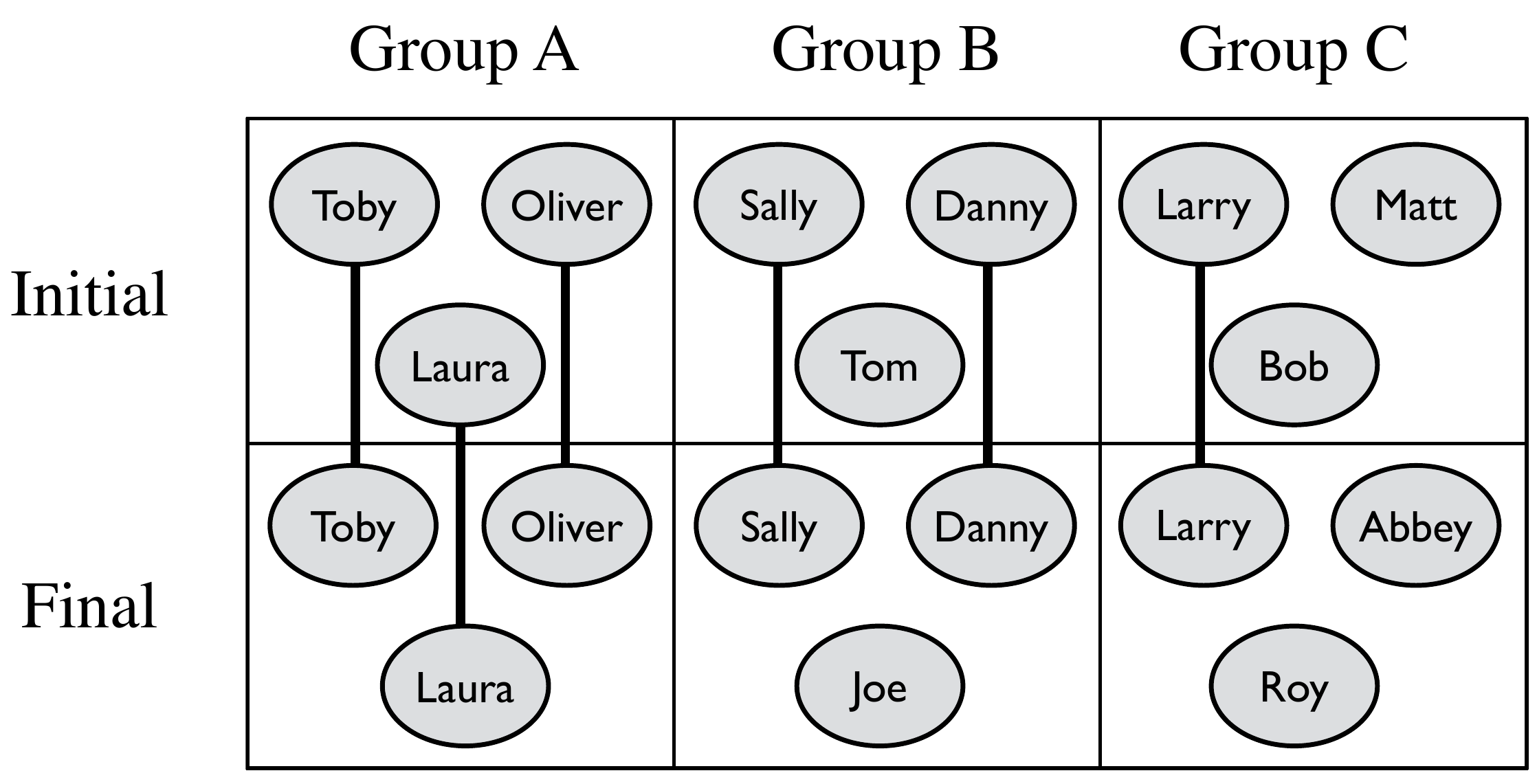} 
\caption{Group membership at the beginning and end of the semester.}{} 
\label{fig:groupmembers} 
\end{center} 
\end{figure}

The primary data for this analysis comes from observations of students participating in AdLab. Lab groups of three students were observed twice a week for three hour class sessions.  This paper focuses on data from the first two weeks of the semester and the last two weeks of the semester.  We follow three separate groups at both times. One of the groups (Group A) remained the same for the whole semester. Group B changed one member at the halfway point. In Group C, only one group member remaining the same. Figure \ref{fig:groupmembers} shows group membership and changes over time\footnote{All names are pseudonyms.}.

The secondary data for this paper comes from semi-structured interviews with students who were recruited from upper-level physics courses in electromagnetism, mechanics, modern laboratory and AdLab as part of an ongoing identity study. Only data from AdLab students are included in this analysis, including interviews from before, during, and after their time in the course. We developed a 45 minute semi-structured interview protocol drawing on identity formation, epistemological sophistication, and metacognition literature that also focused on asking the students to describe their AdLab experiences. The interviews were video-taped and transcribed for analysis.

For supporting evidence, we conducted discussions with the course instructor about her goals for the course and how her instructional choices supported them.  We also collected course artifacts such as instruction manuals and the syllabus.

\subsection{Analysis methods}

Starting with a macro-level of analysis, we looked at each class period and referred to our field notes in order to identify ``activity segments'': all activities whether whole class, particular lab group or individual that occurred during each laboratory session \cite{Kelly1997}. With this index of activities we mapped the events of the classroom over time\cite{Green1981, Kelly1997}. This mapping process allowed for analysis both on a topical level and a sequential level and the identification of thematic content. 

One theme that emerged from our data was that the different student groups doing separate experiments began to talk to each other more frequently and with higher quality interactions as the course progressed.  Another theme that emerged was that both the students and the instructor felt that the physics material and scientific practices in AdLab were closer than previous laboratory classes to ongoing research of practicing physicists. (Other themes emerged; they are not the focus of this paper and will not be discussed here.)  We selected these themes for further study and analysis to help us understand how community of practice develops in the advanced laboratory community and how the AdLab experience affects the professional development of students in the course.

The micro-ethnographic analysis began by first identifying interactions between different groups of students.  The geometry of the AdLab room helped us identify cross-group conversations: while working on a given experiment, a lab group tends to stay clustered around the equipment.  We point the camera at the equipment.  When a student from another group chats with our group of interest, he or she tends to physically visit the group of interest. 

After all of the interactions had been identified, we began to look at the context and content of the inter-group interactions.  We considered the pre- and post-context of the interaction, student discourse (content, tone of voice, volume of speech, and rhythm of turn taking), and body language of the interaction to interpret how the participants frame the interaction. Framing refers to the resources the students bring to bear for a particular interaction\cite{Sayre2008, Irving2013b}. 

Once the different ways of framing inter-group interactions were identified, we then purposefully sampled specific episodes which represented significant evidence of each type of frame. This analysis of interactions with this micro-ethnographic approach allowed for a correlation to how these interactions related to the development of a community of practice within the advanced laboratory classroom. In order to provide further evidence for our claim that a community of practice developed we also quantitatively assessed how the number of interactions between groups changed over time and how the amount of time spent having interactions also changed over time.

As we developed the themes and our observational evidence for it, we triangulated and refined the theme using data from the semi-structured interviews.  Were students aware that a classroom community developed?  We also consulted with the instructor to investigate how her instructional goals might shape the course.

\section{Observations of community development}

The following section focuses on the observational evidence of the community of practice developing in the AdLab learning environment. Through ethnographic analysis of the emerging theme of development of a community of practice we identified the following episodes which highlight either how the structural features or instructor choices helped this community to form. The episodes where also interpreted to show how the community developed over time from its initiation in the first week. These changes are indicated by the change in negotiated norms and discourse that the students use within the AdLab learning environment.

\subsubsection{Episode 1: Typical first experiment interaction (the brief me on the experiment interaction).}

This episode occurs during the first week of the AdLab course during each groups first experiment. It is the second day of Group C (Larry, Bob, Matt) working on the ``E/M Hoag'' experiment. This is an experiment that uses a cathode-ray tube to measure the charge to mass ratio for an electron by sending electrons down a tube with a known magnetic field supplied by a solenoid. The group struggled on the first day to get the experiment successfully set-up to allow for the taking of experimental data but by the time of this episode on the second day they are just at the point where they are successfully taking data. Carl from another group walks by Group C and spots them sitting closely together staring at a screen and decides to ask them how their experiment is going. This episode occurs due to the two of the four structural features being in place: \textit{all in the room together} and \textit{same experiments at different times}.

\begin{description}
\item[Carl:] Whats going on over here? 
\item[Bob:] We're just getting numbers now 
\item[Matt:] (wearily) Lots of numbers
\item[Bob:] (sarcastic tone) Very technical
\item[Carl:] Just looking at that thing, it looks ancient
\item[Bob:] (getting more excited) It's funny sometimes the voltage will drop by hundreds of volts and to fix it you turn it off and turn it back on (makes a ``can you believe this'' face) also this knob broke off so we use a screwdriver to turn it, this knob doesn't even exist anymore.
\item[Carl:] Nice...I think I'll avoid this one
\end{description}

This was a very typical interaction at the start of the semester as students took note of what other experiments different groups where doing and enquired as to the level of difficulty that they involved. The students are aware that they have to do one of the experiments in the room next and because they are \textit{all in the room together} and are doing \textit{same experiments at different times} it allows them the opportunity to discuss the different experiments with their colleagues.

The briefness of this exchange is also typical of the first week of the semester. The AdLab community of practice had not fully negotiated the norm associated with the amount of time these enquires about experiments could last. In the first week these exchanges where all tentative and brief in nature and the students kept to their own group the majority of their time in lab as evidenced by the results in Table 1.1.

Another regularity of the beginning of the semester was the superficial nature in which the Bob talks about the problems with the experiment. His problem is not with the theory behind the experiment or the setting up of the equipment (both of which his group and him had significant trouble with). Instead his focus is on the machinery being dated and problematic. During the first experiment groups would often have intergroup conversations about the difficulty associated with particular experiments but did so superficially. This could be attributed to the development aspect of the bounded community of practice. The norm for how such conversations should occur had not been fully negotiated yet.

This episode indicates the need for the classroom structural features of \textit {all in the room together} and \textit{same experiments at different times} to be present in order for intergroup interactions to occur. These interactions are vital to the development of a community of practice. This episode also indicates that during the first experiment the development process was still occurring and the norms for the community had not yet been negotiated. 

\subsubsection{Episode 2: The ``brief me on the experiment'' interaction in week 8.}

This episode occurs in week 8 when Larry from Group C has now changed groups and is currently working with Abbey and Roy on the Microwave Optics experiment. It is the last day for all groups on their respective experiments and they are all in the process of deciding what experiment to do next. Liam from another group approaches Larry and asks him about the ``E/M Hoag'' experiment which he completed as his first experiment. Essentially this is a repeat of the ``whats going on over here?'' interaction that is described in episode 1. 

Although the types of interactions progressed from just asking how an experiment is, the ``whats going on over here?'' interaction continued regularly but the quality of the interaction increased over time. As before this episode occurs due to the structural features of \textit {all in the room together} and \textit {same experiments at different times} but also \textit {experiments long and hard} as students try to preempt troubleshooting before the experiment begins by asking more detailed questions about the experiment to help with their decision making process.

\begin{description}
\item[Liam:] Did you do that one before? (pointing in the direction of a laboratory bench)
\item[Larry:] The rubidium? (pause) Oh, ``E/M Hoag'', yeah
\item[Liam:] How was that, like for, for theory?
\item[Larry:] (enthusiastically) Basically I combined the theory and derivation, I just talked about, so we've got this device, how can you get a measurement for E over M for the solenoid, you know for the magnetic field and everything, so in talking about how the field was created inside the solenoid and how that affected the path of the electron I felt that covered the theory.
\end{description}

Larry continues to answer several more questions about the experiment before Liam is satisfied with whether he should recommend doing the experiment next to his group.

This episode indicates the change in quality of the intergroup interactions as the community norms have been negotiated at this point in the semester. It is now a large part of the community of practice to have long detailed discussions about the experiments and to that enquiry about specific details of an experiment is okay and revealing specific experiment based expertise to other group members is also okay. The groups are becoming more collaborative. This is evidence of the evolving nature of the community of practice as collaboration becomes more frequent and constructive once the norms of the community have been negotiated.

\subsubsection{Episode 3: Social interactions.}

Episodes 1 and 2 focused on the ``brief me on the experiment'' interaction. This was not the only type of interaction that occurred in the AdLab community of practice. Social interactions were also infrequent to begin with but as with the previous interactions became more prevalent once the community had negotiated its norms that related to what was acceptable as a social interaction in this community. These interactions ranged from the frivolity of cracking jokes to discussions about topics that would be considered off topic but often inspired by some aspect of the experiment they are engaged in. 

In the following episode Matt and Larry are no longer working in the same group but are, for their corresponding experiments, working in close proximity. It is week 9 and Matt has just completed the experiment that Larry is now working on: ``Microwave Optics''. In this experiment students are expected to demonstrate the wave nature of light in a number of interference, diffraction and reflection experiments using microwaves. Matt is currently working on Scanning Tunneling Microscope with his group. This episode demonstrates the camaraderie and social aspect of the community of practice that evolved over time.

\begin{description}
\item[Matt:] (concerned tone) Are the microwaves on?
\item[Larry:] Well they are going this way (indicates the direction he thinks the waves are going).
\item[Matt:] They're reflecting onto your crotch.
\item[Larry:] (laughs) Oh yeah your right, oops, I was like I'll make sure that Percy and Matt are not in the line of fire, I forgot to make sure I wasn't in the line of fire....thanks for your concern about my crotch.
\item[Matt:] (smiling) Your welcome.
\end{description}

This episode can be interpreted in two different ways. Firstly it has an obvious component of Matt playing Larry's set-up of the experimental equipment for humor by referencing the rays reflecting on his crotch. Humor can have a large effect on community building and is a form of discourse that can emphasize membership. The getting of a joke can illustrate that ``you are one of us'', just as missing the humor behind a joke can result in alienation from a community. This is a joke situated within the lab community and presence of such social interactions indicates the development of a community or practice.

The second interpretation is that because of the structural features of the classroom \textit {all in the room together,  experiments long and hard} and \textit {same experiments at different times} this interaction is able to occur. If Matt had not completed the ``Microwave Optics'' lab previously; was not in the room with Larry; had not built up the content expertise and had the time to pay attention to what Larry was doing then he may not have the ability to say anything about Larry's setup. Incidentally there is also an affective element to this interaction as well, Matt is genuinely concerned that Larry is doing something wrong that might have negative effects on Larry in some capacity, even though it is communicated through humor. It indicates an element of the affective nature of communities in that members will look out for their fellow members.

\subsubsection{Episode 4: Experiment Specific Experts.}

This episode focuses on the \textit {experiments long and hard} structural component. Oliver, Toby and Laura are working on the ``Millikan Oil Drop'' experiment. For this experiment the students are attempting to measure the charge on the electron by measuring the charge on small oil droplets and relating this charge to being a multiple of some quantized charge unit. It is week 8 of the AdLab class and this is the groups second day working on the experiment. Toby has been inputting the results the group have been getting so far into his laptop and both he and Oliver are confused about how the equation related to the experiment needs to interpreted with their results and decides to ask for help from Tom who completed the ``Millikan Oil Drop'' experiment the previous week.

\begin{description}
\item[Oliver:] So which is the first plate? Is that the bottom here?
\item[Toby:] Lets ask someone, hey Tom, I have a question for you (Tom walks over)
\item[Tom:] This is d and this is the equation here (points at a point on Toby's screen).
\end{description}
Tom proceeds to spend at least the next 5 minutes explaining his interpretation of their results so far.

As the students began to have a history with experiments the amount of ``can you help me'' interactions increased dramatically as evidenced in both the observational data and in the interviews. The students would make reference to not being able to complete a given experiment if it was not for another group helping them out at a critical juncture. The helping of other students is a clear indication of a community developing with students building up experiment centric expertise and then sharing this expertise due to all of the highlighted structural features of the learning environment but especially \textit {paucity of instructor time} and \textit {experiments long and hard}. At the beginning of the AdLab students would rely on the instructor to help them out and this was often limited to a sort of take a ticket for instructor time set-up. By the end of the AdLab students who had now developed experiment centric expertise where now being asked for help and would freely oblige often spending upwards of 30 minutes helping another group.

\subsubsection{Episode 5: AdLab Based Discourse.}

As mentioned in episode 3 and in the section on community of practice a big part of being integrated into a community is to begin to appropriate the discourse of the community. If the misinterpretation of jokes can lead to alienation so can the inability to communicate in the language of the community. The following episode occurs as Sally, Danny and Mike are on their second day of working with the nuclear magnetic resonance (NMR) spectrometer.  The NMR has multiple possible experiments designed for  use with the equipment, some of which are reliant on obtaining the Free Induction Decays (FID) signal on an oscilloscope. Oliver, Toby, and Laura had previously completed the NMR experiment and are working on the ``Millikan Oil Drop'' experiment, which is not located next to (but is within sight of) the NMR set-up. Laura had just borrowed an ruler off Sally. While she is returning it, Laura relays a message from Oliver to Sally's group. 

\begin{description}
\item[Laura:] Thanks Sally, Oliver says nice FID signal
\item[Sally:] (laughs) thanks
\end{description}

Although brief, this example gives a great sense of the development of the AdLab community of practice as by this time period of the community, experiment specific discourse has become ubiquitous amongst those who have carried out certain experiments. It wasn't just nice signal, it was nice ``FID'' signal. The students began to develop and appropriate the language of the community and use it within the classroom. 

Another element of community development that is in evidence in this exchange is the fact that Oliver feels comfortable to comment on another groups experiment and how well they are doing. The groups moved from a beginning point where they where insulated groups occasionally discussing how hard an experiment was to the point where they are freely discussing, socializing and evaluating each others work on a regular basis.

\subsubsection{Episode 6: Instructional Choice.}

The final episode focuses on the other crucial element present for the AdLab community of practice to develop and that is the instructional choice of the instructor. Toby, Oliver, and Laura are working on the ``Millikan Oil Drop'' Experiment as described in episode 4. It is the first session of the new experiment and the instructor comes over to quiz them on how their first tentative steps to setting up the experiment is going. In the ``Millikan Oil Drop'' experiment there is a choice of several oil atomizers that can be employed in the setup of the experiment and this is the focal point of the initial discussions. The instructor due to the structural constraints of the lab at this point in the semester is not aware of which atomizer has been working best and invites over a member of the previous group that has carried out the experiment (Tom) to discuss with Toby, Oliver, and Laura the expertise he has developed.

\begin{description}
\item[Instructor:] Did you guys do this last?
\item[Tom:] My group did it last (volunteering quickly from other end of room)
\item[Instructor:] Good, do you have any tips for them?
\item[Tom:] (Tom walks over) Um, get used to taking it apart and cleaning it
\item[Instructor:] Okay, keep cleaning it a lot
\item[Tom:] Yeah do that a lot, if you get a build up, if there is a big white blotch, the top which is actually the bottom of the T.V. screen 
\item[Instructor:] It's labeled top but it says bottom because its inverted right?
\item[Tom:] Yeah, if you get a big blotch there you can probably, its a build up of oil, you have a little thing to dab it out, dab and then dab it on a paper towel
\item[Instructor:] Use this to dab it out?
\item[Tom:] Yeah
\item[Instructor:] Oh thats nice, so you don't have to take it all apart?
\item[Tom:] Yeah, yeah
\item[Instructor:] Okay
\item[Tom:] That's an easier way of getting rid of some of the excessive stuff
\item[Instructor:] Okay thats a good tip
\item[Tom:] Em, thats about it
\item[Instructor:] Okay
\item[Tom:] (enthusiastically) It's fun if you get it to work
\end{description}

By inviting Tom over the instructor is sublimely negotiating the norm that it is okay to consult with other groups especially those who have previously completed the concerned experiment for help and advice. This encouragement of the development and sharing of knowledge and resources is a deliberate choice by the instructor due to the structural features of AdLab learning environment. 

Episodes 1-6 have been presented above to demonstrate how the structural features and a instructional choice on behalf of the instructor encouraged intergroup cooperation and collaboration that has helped to develop a classroom community in the AdLab course with enculturation elements. The episodes emphasize the importance of there structural features and how they are connected to the development of specific elements of our classroom community of practice like the negotiation of norms or distributed expertise . The above episodes are a tiny minority of episodes that could have been chosen as evidence of the development of a learning community with these enculturation elements. In the next section we present quantitative evidence of how often groups interacted as further evidence of the many central participants and level of community involvement and collaboration elements of the learning community.

\subsection{Quantitative analysis of community talk}

 \begin{table} 	
 \caption{Inter-group interactions at the beginning and end of the semester.  Numbers are percent of total time spent talking to other groups in the second experiment of the semester (Initial) and penultimate lab of the semester (Final).} 	
 \centering
 \begin{tabular}{|c|c|c|c|}
 \hline
Time Period of Semester&Group A&Group B&Group C\\
\hline
Initial & 1.8\% & 0.9\% & 5.1\% \\
\hline
Final & 12.4\% & 8.0\% & 17.3\% \\
\hline		
 \end{tabular}
 \label{tab:intergroup} 
 \end{table}

Table \ref{tab:intergroup} presents the percentage of laboratory time that the three groups observed spent interacting with another group in the laboratory environment at two different time periods. ``Initial'' refers to each groups percentage interactions with other groups during their second experiment of the semester. An experiment typically lasted 4 classroom sessions over a two week period which would be approximately 12 hours class time. The ``final'' is the percentage interactions with other groups for their second-to-last experiment of the semester. By the penultimate experiment of the semester group A has remained static in its membership while group B and C have changed members after their third experiment as indicated in Figure \ref{fig:groupmembers}  An intergroup interaction was coded in one of three ways. The first was if a member of another group came over to the group being observed and interacted with them. The second was if a member of the group being observed left that group to go interact with another group. The third was if groups initiated a conversation or joined a conversation with another group while being physically adjacent to their experimental set-up. The total time spent interacting with other groups by the 3 previously described methods was combined to calculate the total amount of time spent interacting with other groups.

The results indicate that the difference in time spent interacting with other groups between the two time periods ``initial'' and ``final'' for all three groups is significantly different. The amount of interactions that each group had with other groups at the start of the semester are substantially less than the amount of interactions at the end of the semester. The consistency in the difference between amount time interacting with other groups between the ``initial'' and ``final'' time periods across all groups allows a claim that more classroom discourse was occurring between groups by the end of the semester. This is compelling argument that a community of practice did develop over time in the advanced laboratory community. There are differences between the increase in interactions between groups especially in the case of group B which as a group did not increase in the amount they interacted with other groups as greatly as the other two groups. Although group membership and personality may account for the difference it is worth noting that group B's final experiment was the NMR set-up. The NMR experiment was new to the advanced laboratory learning environment and the groups that had completed the experiment prior to group B all struggled with it. This resulted in the instructor spending more time with the group than was typical and preempting problems that the group may have sought solutions for from prior groups. Overall though these results demonstrate that all 3 groups became further involved in the community as the semester progressed.

\section{Interview reflections on community development}

As part of the longitudinal study examining how upper-level physics students develop an identity as  a physicist we conducted semi-structured interviews on a regular basis for the majority of this group of students. One of these sets of interviews was conducted at the 10 week point into the AdLab semester. As part of this interview we enquired about the students experiences in the AdLab. An important theme to emerge from the interview data is that the students also noticed several of the structural features that promote community development. In the following sections we discuss extracts from the interviews that pertain to specific structural or community building factors.

\subsubsection{Extract 1: Paucity if instructor time.} 

In extract 1 the interviewer asks Matt what he thought of the approach to instruction that was taken in the AdLab environment. From observations by the investigators in AdLab sessions they noticed that in the beginning of the semester there was often a queue for the instructors attention but that this became a less prominent feature of the classroom as time passed. We wanted to know if the students were aware of this and how did they feel about a perhaps perceived lack of access to the instructor. 

\begin{description}
\item[Matt:] It was pretty well taught but there was a lot of people in there so we couldn't get a lot of one on one time, when we needed help...so two of our experiments, the first two, we where the first group doing so we couldn't ask anyone else about them, but the other ones, when we couldn't consult with (Instructor) we went to the people that had already done that experiment and they were usually able to figure what it is we were missing or what went wrong when we were setting up like that.
\end{description}

Matt specifically references the amount of people in the room and the lack of one on one time when help was required. This is Matt noticing the structural feature of \textit{paucity of instructor time} and indicating that this was something he found problematic at first. This was resolved once the other groups in the lab and himself had build up experiment specific expertise and began to consult with each other. The consulting with each other and experiment specific expertise are further evidence of the structural features \textit{all in the room together, experiments long and hard} and \textit {same experiments at different times} although Matt is not being as explicit about the last 3 features.

\subsubsection{Extract 2: Experiments long and hard}

Extract 2 covers all four of the structural features again but Toby's reflections refer more explicitly to the \textit{experiments long and hard} aspect of the structure. Toby is answering the same question as Matt did in extract 1 in regards to what he thought of the approach to instruction and describes spending time working with other students. The interviewer follows up by asking Toby specifically about collaboration and working with other students.

\begin{description}
\item[Int:] How did you collaborate with the other people and what did you get from the other people in advanced lab?
\item[Toby:] If we ever had a problem, like we had a problem, with the Zeeman experiment. We couldn't quite figure out how we where supposed to set it up, so we went to Mike, asked him and he showed us how he did it. For NMR (referring to the group currently doing that experiment), they weren't quite sure what they where doing so they had Oliver and me come over. Mainly Oliver but I helped a little bit. We did the ``E/M Hoag'' [experiment]. For the ``E/M Hoag'' we had to derive the equation we needed and we went to eh Larry and Roy and we were able to look at their work and see what they did and once we saw where they started it wasn't particularly hard to get it. So we basically drew on their experience, everyone seemed to draw on the experience of the experiments everyone else had when starting.
\end{description}

Toby's description of the give and take of assistance between groups over several experiments indicates the growth of a community of practice. The \textit {paucity of instructor time} is referenced in Toby's description of going to another group when a problem arose as opposed to the instructor. The \textit {experiments long and hard} is indicated by Toby seeking out other groups to help with equipment set or derivations and the other groups had both the expertise and the time to help them out and reciprocally Toby had the time to help other groups when they had similar problems. Doing the \textit {same experiments at different times} allows the experiment specific expertise to develop.

\subsubsection{Extract 3: Community Development}

A portion of each interview was aimed at examining how students perceived what they where getting out of their advanced laboratory experiences. For the most part this involved students describing how the experience had helped them understand the material but some questions where directed at asking what they thought particular elements of the course design where for. In extract 3 the interviewer asks Tom what he thought the purpose of the presentations that each student had to perform once a semester where for.

\begin{description}
\item[Int:] So what do you think the point is behind the presentations?
\item[Tom:] So we have to present things in real life, we have to talk to people....it also strengthens our knowledge of the experiments and builds a community in the class, you get to talk to other people.
\end{description}

Tom thinks that the presentations are a part of the course in order to foster real world experiences or in other words develop some authentic physicist practices. Students identifying aspects of the course that they perceived as contributing to their preparedness for future endeavors in the interviews was common. It was also common that students made reference to collaborating or working with other groups as indicated in extract 2 and 3 as Tom does by identifying explicitly that the goal of the presentation activity is community development driven. 

\subsubsection{Extract 4: Development over time}

As with extracts 1 and 2 part of the semi-structured interview focused on collaboration with other groups and students. In the description of bounded communities of practice earlier in the paper we described that they did not just occur when you put a group of people together in a room. A development process has to occur and norms have to be negotiated. In extract 4, Tom reflects that he did not ask other groups about labs in the beginning but that this changed over time.

\begin{description}
\item[Int:] So did you ask other people about labs often?
\item[Tom:] At the start not really. I kind of just kept to my group, except, well with the other groups that I knew I made jokes with, I'd hear things and just make jokes. I'm doing it more now, other people are talking to me as well about labs
\end{description}

It was indicated in section 5 that this process of isolated groups becoming more interactive over time was observed both in the quantitative and qualitative observational results. Tom reflecting on the process is further evidence that the community developed over time. The next section will also reflect on interview data but will focus on the other element of the AdLab being classified as a crucible course: the steep rise in ADK.

All of the above extracts provide further evidence that the enculturation elements developed within the classroom community over time and that students where aware of some of these elements.

\subsection{Students' descriptions of AdLab as a jump in accountable disciplinary knowledge}

Another feature of the AdLab learning environment is that there is a quite observable jump  in accountable disciplinary knowledge\cite{Stevens2008} (ADK) from the previous courses that the students would have taken. Accountable disciplinary knowledge is ``what counts'' as doing physics: the kinds of activities, problems, and discourse that people engage in when they are participating in a physics community of practice.  For example, doing well at the introductory physics level often entails solving 15 end-of-chapter problems weekly in a few hours alone and doing well at the upper-division undergraduate level entails solving a few problems weekly in 15 hours with peers. This difference in ``what counts'' as doing physics well constitutes a jump in accountable disciplinary knowledge between introductory and upper-division physics.  Evidence of an ADK jump in AdLab is very striking in students' descriptions of the course after participating in it for one semester. 

\begin{description}

\item[Tom:] The labs are more complex and more interesting. A lot less hand holding. There more enjoyable and they are actually looking at phenomena that I am interested in\dots{}its more about us discovering the phenomena\dots{}it feels like more of a professional setting than most of my other courses.

\item[Matt:] we have been investigating actual atomic structures or how to find the mass of an electron\dots{}previous labs would be a lot more cut and dry. Here's the procedure. Follow it. You'll get the results, easily, these ones where more of, heres the procedure. Most of it usually. Follow it and try and understand whats going on cause if you don't you won't know if what your getting is any good\dots{}the real feeling of being a physicist was trying to understand all that stuff that we get from it.

\item[Laura:] I really had to do a lot of work on my own and I wasn't really expecting that\dots{}I thought maybe the lab write ups would be a little bit more prescribed and not so quite, its kind of like, these are your objectives, this is how the machine works, do it, and thats good.

\item[Toby:] Yeah the subject matter itself changed but thats to be expected for a higher level class\dots{}obviously they are trying to get you to really think about the subject matter. To understand the subject matter at a deeper level than just in EP labs. They want you to see it happen in advanced lab. They want you to see it happen and understand why its happening, by figuring it out yourself rather than being told. I mean we don't want to create people who can just rattle of equations without understanding what those equations really mean. You want people who actually understand what those equations really mean\dots{}this time we have a lot more freedom in the time that it takes to do it. You know we have some constraints because the other groups have to use the equipment as well but we can come in on our own and do it. The freedom was nice even if it was the result of having more work.

\end{description}

Several of the students perceive that a lot of what they were doing in the AdLab environment and   how they participated in it where more like authentic practices of physicists. The students also clearly perceived a jump in the level of the material and what was expected of them in the AdLab classroom. Changes in expectations are obvious from students noting that there was a lot more freedom and the labs where more prescribed and that they were expected to gain an understanding of the material and not just get a set of data.

\section{Relating Communities of Practice to Accountable Disciplinary Knowledge: Crucible Courses}

The combination of quantitative and qualitative results presented in this paper clearly indicate that a classroom community of practice developed in the AdLab learning environment with certain enculturation elements. It is also obvious from the students reflections on the course that there was was a significant jump in ADK from their previous experiences due to significant changes in the structural and programmatic features of the AdLab learning environment. This combination of jump in ADK and enculturation elements of the course resulted in students being offered the opportunity to accelerate their own trajectory to being more of a central participant of the physicists community of practice. This emphasis on enculturation from both ADK and structure has resulted us in labeling the AdLab course as a possible ``crucible course''.

We describe crucible courses as the first courses in which students work on difficult physics problems surrounded primarily by other physics students, are treated by their professors as junior physicists, and take on identities as part of a community of physics students. In our prior work, we have identified crucible courses: those associated with large changes in ADK and developments of physics identity.  Both students \cite{Irving2013} and researchers\cite{Sayre2008} seem to know these courses ``when they see them'' \cite{Jacobellis1964}. The courses are typically intermediate level (taken by sophomores or juniors) and are among the first courses populated predominately by physics majors and minors. They have smaller enrollments and foster a greater sense of community within the class.  They may be theory courses or laboratory courses, but in either case the expectations of students and their perceptions of the stakes are substantially higher than in previous courses. This is a working definition and we intend to further investigate what are the key elements of a crucible course.  We  believe an emphasis on enculturation is a key feature of a crucible course.

To discuss further this enculturation process we must first examine communities of practice and how we interpret they fit into the college environment. In alignment with previous researchers \cite{Schoenfeld1992, Duschl2002, Demaree2009, Lemke1990, Berland2011} we believe it is applicable to view the classroom community as a community of practice. If that is the case as a student you will occupy many communities of practice concurrently within the college environment while also being a member of several other communities outside of the college context. In fact the majority of students waking hours during their time in college will not be spent in the classroom \cite{Li2011}. The combination of these memberships to a variety of communities of practice will all have influences on each other and can help in the development of a physics identity both in obvious and less obvious ways. 

Students are on trajectories to developing an identity as a physicist when they enter a physics classroom. Once they enter a physics classroom they are developing a relationship with physics that may turn into a physics identity. They may not intend on becoming a physicist, it might not even be their major but when they enter a physics classroom they engage in a variation of the practices of becoming a physicist. That is the nature of the a classroom being a community of practice and so in essence any physics classroom is a sub-community of the community of physicists.

\section{Discussion}
All classroom communities of practice are different and these differences may be trivial or may be extensive. Different classroom communities offer different levels of exposure to the authentic expectations, practices, content knowledge and discourses of the discipline of physics. Therefore the differences between classroom communities of practice can result in students moving along their trajectories to being a member of the physicists community of practice at different accelerations. To clarify the classrooms would offer the opportunity but it is up to the students to participate either peripherally or centrally.

The AdLab classroom is an example of a community of practice that offers the opportunity to have an accelerated trajectory towards being a central participant of the community of practice of physicists. It introduces students to the expectations, practices, content knowledge and discourses that more closely resemble those of the physicists community. This is achieved by having the students collaborating as a group and with other groups on long and hard physics experiments that are generally more modern in setting in an environment that echoes what students might perceive as a research environment. Being a central participant in this environment will accelerate ones trajectory to being a more central participant of the community or practice of physicists. 

We do not think that all classroom communities of practice should offer opportunities of accelerated trajectories. An accelerated trajectory classroom in introductory physics would be inappropriate. It has been indicated \cite{Borasi1998} that there are already great shifts being expected of students in introductory classes as teachers try to move students away from being socialized to memorize, practice and recite and move towards being comfortable with constructivist and social constructivist perspectives. Also the norms of college can be very different from the norms of school and again the norms of actual practitioners of physics. 

We have argued that AdLab develops into this a community of practice very effectively due the factors of: paucity of instructor time; all in the room together; experiments long and hard, do some of the same experiments at different times; instructor supports the development of a community of practice. Of the above claims all of them have been discussed extensively in the results except for \textit {All in the room together}. This claim comes from the assertion that in the previous semester some of the same students took the modern laboratory course which is set up so that each group of students attempted the same experiment each week and so no community of practice could develop except amongst each separate group of 3 or 2 students. When asked in interviews whether they had discussed the laboratory they where trying to complete with other members of the modern laboratory class the answer was generally no, although they often did work with their group outside of class.

As mentioned previously the development of a community in community of practice literature is not commonly discussed but to us is a key feature of bounded communities of learning. A community of practice in a classroom does not form on the condition of putting students in a room together although it may result in one eventually. In our case a classroom community of practice developed that had several elements: classroom norms and expectations; distributed expertise; community involvement; many central players and the instructor is not sole mediator. We argue that these elements developed as a result of the presence of all structural features of the classroom previously mentioned and the instructors choice to emphasize collaboration. We also argue that these elements are an important part of the AdLab course offering the opportunity for a student to accelerate ones trajectory towards being a more central participant of the physicists community of practice.

The classroom community did not start with the enculturation elements. Students' ways of participating change as they learn the norms and practices of the classroom community of practice, which includes developing a shared discourse with their fellow community members of students and instructors \cite{Bielaczyc1999}. The students also have to figure out the boundary constraints \cite{Barab2002} of this new community of practice due to it being a bounded community. Norming is one of the five stages of group development\cite{Tuckman1965} and although not necessarily relatable to the communities of practice theory does indicate that a classroom has to go through some development before it becomes the finalized version of the learning community. We believe for the AdLab classroom the features previously mentioned are the reason why it developed into a classroom community with several enculturation elements and the majority (if not all) of the students participating centrally. 

A big jump in ADK from course to course can be difficult for students as often what they think doing physics means has changed from what it has meant in the past. It could be argued that the community of practice developing is a support mechanism for the students in order to deal with the jump in ADK. A big jump in ADK without a community developing could result in greater losses in retention and persistence as students struggle to deal with the changes in norms and expectations. Added to this is that cultural practices of professional scientists are always adapted to fit the realities of the classroom and to suit the teachers values/goals \cite{Hogan2001b, Squire2003}. When designing curricula or courses careful consideration should be given to the expectations, practices, content knowledge and discourses of the community of physicists that are being incorporated into the design. A realization must be made that what we ask of the students is not just different content but a different level of content that attached to it has a different set of norms and expectations. 

Adding structural and instructional features to a course that encourages the development of an effective community of practice may be one way of equipping students to deal with such transitions. 

\section{Conclusion} 

The advanced laboratory community of practice was identified as a community of practice that provides the opportunity to accelerate a students trajectory to becoming a member of the physicists community of practice. It was also identified as a community of practice that forms quickly and has several enculturation elements due to several reasons: paucity of instructor time; all in the room together; experiments long and hard, do some of the same experiments at different times; instructor encourages the sharing and co-development of knowledge and understanding.

\bibliographystyle{aipproc}

\begin{thebibliography}{55}
\expandafter\ifx\csname natexlab\endcsname\relax\def\natexlab#1{#1}\fi
\providecommand{\enquote}[1]{``#1''}
\expandafter\ifx\csname url\endcsname\relax
  \def\url#1{\texttt{#1}}\fi
\expandafter\ifx\csname urlprefix\endcsname\relax\def\urlprefix{URL }\fi
\providecommand{\eprint}[2][]{\url{#2}}

\bibitem[Pierrakos et~al.(2009)]{Pierrakos2009}
O.~Pierrakos, T.~K. Beam, J.~Constantz, A.~Johri, and R.~Anderson, \enquote{{On
  the development of a professional identity: engineering persistors vs.
  engineering switchers.},} in \emph{Frontiers in Education Conference}, IEEE,
  2009, vol.~39, pp. 599--604.

\bibitem[Barton and Yang(2000)]{Barton2000}
C.~A. Barton, and K.~Yang, \emph{Journal of Research in Science Teaching}
  \textbf{37}, 871--889 (2000).

\bibitem[Chinn and Malhotra(2002)]{Chinn2002}
C.~A. Chinn, and B.~A. Malhotra, \emph{Science Education} \textbf{86}, 175--218
  (2002).

\bibitem[Cleaves(2005)]{Cleaves2005}
A.~Cleaves, \emph{International Journal of Science Education} \textbf{27},
  471--486 (2005).

\bibitem[Shanahan(2007)]{Shanahan2007}
M.~C. Shanahan, {Playing the role of a science student: Exploring factors and
  patterns in science student identity formation.} (2007).

\bibitem[Hunter et~al.(2006)]{Hunter2006}
A.-b. B.~A. Hunter, S.~L.~S. Laursen, and E.~Seymour, \emph{Science Education}
  (2006),
 
\bibitem[Bonnar(2007)]{Bonnar2007}
I.~D. Bonnar, \emph{{Not as cool as fighter pilots}}, Ph.D. thesis, University
  of Stirling (2007).

\bibitem[Lave and Wenger(1991)]{Lave1991}
J.~Lave, and E.~Wenger, \emph{{Situated learning: legitimate peripheral
  participation.}}, Cambridge, UK: Cambridge University Press., 1991.

\bibitem[Del-Castillo et~al.(2003)]{Del-Castillo2003a}
H.~Del-Castillo, A.~{Bel\'{e}n Garc\'{\i}a-Varela}, P.~Lacasa, and A.~B.
  Garcia-Varela, \emph{International Journal of Educational Research}
  \textbf{39}, 885--891 (2003), ISSN 08830355,
  \urlprefix\url{http://linkinghub.elsevier.com/retrieve/pii/S0883035504001004}.

\bibitem[Li(2011)]{Li2011}
S.~Li, \emph{{Learning in a Physics Classroom Community: Physics Learning
  Identity Construct Development, Measurement and Validation}}, Ph.D. thesis
  (2011).

\bibitem[Danielsson(2007)]{Danielsson2007}
A.~T. Danielsson, \emph{{The gendered doing of physics: A conceptual framework
  and its application for exploring undergraduate physics students' identity
  formation in relation to laboratory work}}, Ph.D. thesis, Uppsala Universitet
  (2007).

\bibitem[Wenger(1998)]{Wenger1998}
E.~Wenger, \emph{{Communities of practice: learning, meaning, and identity}},
  Cambridge University Press, 1998.

\bibitem[Coakes and Clark(2006)]{Coakes2006}
E.~Coakes, and S.~Clark, \emph{{Encyclopedia of Communities of Practice in
  Information and Knowledge Management}}, Hershey, PA: Idea Group Reference,
  2006, chap. "The Conce.

\bibitem[Barab et~al.(2002)]{Barab2002}
S.~Barab, M.~Barnett, and K.~Squire, \emph{The Journal of the Learning
  Sciences} \textbf{11}, 489--542 (2002).

\bibitem[Nespor(1994)]{Nespor1994}
J.~Nespor, \emph{{Knowledge in motion: Space, time and curriculum in
  undergraduate physics and management}}, Falmer Press, Washington, D.C., 1994.

\bibitem[Charney et~al.(2007)]{Charney2007}
J.~Charney, C.~Hmelo-Silver, W.~Sofer, L.~Neigeborn, S.~Coletta, and
  M.~Nemeroff, \emph{International Journal of Science Education} \textbf{29},
  195--213 (2007).

\bibitem[Hsu and Roth(2010)]{Hsu2010}
P.~L. Hsu, and W.~M. Roth, \emph{Research in Science Education} \textbf{40},
  291--311 (2010).

\bibitem[Lopatto(2009)]{Lopatto2009}
D.~Lopatto, \emph{{Science in Solution: The Impact of Undergraduate Research on
  Student Learning}}, Research Corporation for Science Advancement, 2009, 1
  edn.
  
\bibitem[Furman and {Calabrese Barton}(2006)]{Furman2006}
M.~Furman, and A.~{Calabrese Barton}, \emph{Journal of Research in Science
  Teaching} \textbf{43}, 667--694 (2006).

\bibitem[Aschbacher et~al.(2010)]{Aschbacher2010}
P.~R. Aschbacher, E.~Li, and E.~J. Roth, \emph{Journal of Research in Science
  Teaching} \textbf{47}, 564--582 (2010).

\bibitem[Aikenhead(1996)]{Aikenhead1996}
G.~S. Aikenhead, \emph{Studies in Science Education} \textbf{27}, 1--52 (1996).

\bibitem[Costa(79)]{Costa1995}
V.~B. Costa, \emph{Science Education} \textbf{3} (79).

\bibitem[Borasi et~al.(1998)]{Borasi1998}
R.~Borasi, M.~Siegel, J.~Fonzi, and C.~F. Smith, \emph{Journal for Research in
  Mathematics Education} \textbf{29}, 275--305 (1998).

\bibitem[Schoenfeld(1992)]{Schoenfeld1992}
A.~H. Schoenfeld, {Learning to Think Mathematically: Problem Solving,
  metacognition, and Sense-Making in Mathematics (ch. 15)} (1992),
  \urlprefix\url{http://gse.berkeley.edu/faculty/AHSchoenfeld/LearningToThink/Learning\_to\_think\_Math.html}.

\bibitem[Lemke(1990)]{Lemke1990}
J.~L. Lemke, {Talking Science: Language, Learning, and Values} (1990).

\bibitem[Berland(2011)]{Berland2011}
L.~K.~L. Berland, \emph{Journal of the Learning Sciences} \textbf{20}, 625--664
  (2011), ISSN 1050-8406,
  \urlprefix\url{http://www.tandfonline.com/doi/abs/10.1080/10508406.2011.591718}.

\bibitem[Nathan(2005)]{Nathan2005}
R.~Nathan, \emph{{My Freshman Year; What a Professor Learned By Becoming a
  Student}}, New York: Cornell University Press, 2005.

\bibitem[Cobb et~al.(2001)]{Cobb2001}
P.~Cobb, M.~Stephan, K.~McClain, and K.~Gravemeijer, \emph{The Journal of the
  Learning Sciences} \textbf{70}, 113--163 (2001).

\bibitem[Rogoff et~al.(1996)]{Rogoff1996}
B.~Rogoff, E.~Matusov, and C.~White, \emph{{The handbook of education and human
  development: New models of learning, teaching and schooling.}}, Cambridge,
  MA: Blackwell, 1996, chap. Models of, pp. 6--42.

\bibitem[Goertzen et~al.(2011)]{Goertzen2011}
R.~M. Goertzen, E.~Brewe, and L.~Kramer, \enquote{{Transforming Participation:
  A Case Study of an Introductory Physics Student in a Modeling Instruction
  Class},} in \emph{COERC 2011 Proceedings of The Tenth Annual College of
  Education and Graduate Network Research Conference}, 2011.

\bibitem[Lampert(1990)]{Lampert1990}
M.~Lampert, \emph{American Educational Research Journal} \textbf{27}, 29--63
  (1990).

\bibitem[Bruffee(1997)]{Bruffee1993}
K.~A. Bruffee, \emph{{Collaborative learning: Higher education,
  interdependence, and the authority of knowledge.}}, Baltimore, MD: Johns
  Hopkins University Press., 1997.

\bibitem[Cockrell et~al.(2000)]{Cockrell2000}
K.~S. Cockrell, J.~A. Hughes-Caplow, and J.~F. Donaldson, \emph{The Review of
  Higher Education} \textbf{23}, 347--363 (2000).

\bibitem[Duschl and Osborne(2002)]{Duschl2002}
R.~A. Duschl, and J.~Osborne, \emph{Studies in Science Education} \textbf{38},
  39--72 (2002).

\bibitem[Squire et~al.(87)]{Squire2003}
K.~Squire, J.~G. MaKinister, M.~Barnett, A.~L. Luehmann, and S.~Barab,
  \emph{Science Education}  (87).

\bibitem[Hogan and Corey(32)]{Hogan2001}
K.~Hogan, and C.~Corey, \emph{Anthropology \& Education Quarterly} \textbf{2}
  (32).

\bibitem[Demaree and Li(2009)]{Demaree2009}
D.~Demaree, and S.~Li, \enquote{{Promoting productive communities of practice:
  An instructor’s perspective},} in \emph{2009 PER Conference Proceedings -
  AIP Conference Proceedings 1179}, edited by M.~Sabella, and C.~Singh, Physics
  Education Research Conferences, American Association of Physics Teachers,
  American Institute of Physics, Melville, NY, Ann Arbor, MI, 2009, pp.
  125--128, \urlprefix\url{http://dx.doi.org/10.1063/1.3266694}.

\bibitem[Pirie(1997)]{Pirie1997b}
S.~Pirie, \emph{Journal for Research in Mathematics Education} \textbf{9}
  (1997).

\bibitem[Gar(1967)]{Garfinkel1967}
\emph{{Studies in ethnomethodology.}}, Hemel Hempstead: Prentice Hall., 1967.

\bibitem[Mar(1998)]{Marcus1998}
\emph{{Ethnography through thick and thin.}}, Princeton University Press, 1998.

\bibitem[Brown(2004)]{Brown2004}
B.~a. Brown, \emph{Journal of Research in Science Teaching} \textbf{41},
  810--834 (2004), ISSN 0022-4308,
  \urlprefix\url{http://doi.wiley.com/10.1002/tea.20228}.

\bibitem[Collins et~al.(2004)]{Collins2004}
A.~A. Collins, D.~Joseph, and K.~Bielaczyc, \emph{Journal of the Learning
  Sciences} \textbf{13}, 15--42 (2004),
  \urlprefix\url{http://www.tandfonline.com/doi/abs/10.1207/s15327809jls1301\_2}.

\bibitem[Ernest(1997)]{Ernest1997}
P.~Ernest, \emph{{Social constructivism as a philosophy of mathematics.}},
  University of New York Press, 1997.

\bibitem[Emerson et~al.(1993)]{Emerson1993}
R.~M. Emerson, {Fretz R. I.}, and L.~L. Shaw, \emph{{Writing ethnographic
  fieldnotes.}}, University Of Chicago Press, 1993.

\bibitem[Case and Light(2011)]{Case2011}
J.~M. Case, and G.~Light, \emph{Journal of Engineering Education} \textbf{100},
  186--210 (2011).

\bibitem[Kelly and Crawford(1997)]{Kelly1997}
G.~J. Kelly, and T.~Crawford, \emph{Science Education} \textbf{81}, 533--559
  (1997).

\bibitem[Green and Wallat(1981)]{Green1981}
J.~Green, and C.~Wallat, \emph{{Ethnography and language in educational
  settings}}, Norwood, NJ: Ablex, 1981, chap. Mapping in.

\bibitem[Sayre and Wittmann(2008)]{Sayre2008}
E.~C. Sayre, and M.~C. Wittmann, \emph{Physical Review Special Topics - Physics
  Education Research} \textbf{4}, 20105 (2008), ISSN 1554-9178,
  \urlprefix\url{http://link.aps.org/doi/10.1103/PhysRevSTPER.4.020105
  http://link.aps.org/abstract/PRSTPER/v4/i2/e020105}.

\bibitem[Irving et~al.(2013)]{Irving2013b}
P.~W. Irving, M.~S. Martinuk, and E.~C. Sayre, \emph{Physical Review Special
  Topics - Physics Education Research} \textbf{9}, 010111 (2013), ISSN
  1554-9178,
  \urlprefix\url{http://link.aps.org/doi/10.1103/PhysRevSTPER.9.010111}.

\bibitem[Stevens et~al.(2008)]{Stevens2008}
R.~Stevens, K.~O'Connor, L.~Garrison, A.~Jocuns, and D.~M. Amos, \emph{Journal
  of Engineering Education} \textbf{97}, 355--368 (2008),
  \urlprefix\url{http://www.engr.washington.edu/caee/CAEE\_Briefs\_PDFs/BecominganEngineering\_Stevens\_JEE08.pdf}.

\bibitem[Irving and Sayre(2013)]{Irving2013}
P.~W. Irving, and E.~C. Sayre, \emph{Journal of the Scholarship of Teaching and
  Learning}  (2013).

\bibitem[Stewart(1964)]{Jacobellis1964}
P.~Stewart, \emph{US Rep} \textbf{378}, 184 (1964).

\bibitem[Bielaczyc and Collins(1999)]{Bielaczyc1999}
K.~Bielaczyc, and A.~Collins, \emph{{Instructional-design theories and models:
  A new paradigm of instructional theory}}, Mahwah, NJ: Lawrence Erlbaum, 1999,
  chap. Learning c, pp. 269--292.

\bibitem[Tuckman(1965)]{Tuckman1965}
B.~W. Tuckman, \emph{Psychological Bulletin} \textbf{63}, 384--399 (1965), ISSN
  0033-2909, \urlprefix\url{http://www.ncbi.nlm.nih.gov/pubmed/14314073}.

\bibitem[Hogan and Maglienti(2001)]{Hogan2001b}
K.~Hogan, and M.~Maglienti, \emph{Journal for Research in Science Teaching}
  \textbf{38}, 663--687 (2001).

\end{thebibliography}
%

\end{document}